\documentclass[aps,superscriptaddress,showpacs,floatfix]{revtex4}
\usepackage{graphicx}
\usepackage{dcolumn}
\usepackage{bm}
\usepackage{CJK}
\usepackage[colorlinks,
            linkcolor=red,
            anchorcolor=blue,
            citecolor=green
            ]{hyperref}

\begin{document}

\title{Sub Coulomb barrier d+$^{208}$Pb scattering in the time-dependent basis function approach}
\author{Peng Yin}
\affiliation{Department of Physics and Astronomy, Iowa State University, Ames, IA 50011, USA}
\affiliation{Institute of Modern Physics, Chinese Academy of
Sciences, Lanzhou 730000, China}
\author{Weijie Du\footnote{Corresponding author: duweigy@gmail.com}}
\affiliation{Department of Physics and Astronomy, Iowa State University, Ames, IA 50011, USA}
\author{Wei Zuo}
\affiliation{Institute of Modern Physics, Chinese Academy of
Sciences, Lanzhou 730000, China}
\affiliation{School of Nuclear Science and Technology, University of Chinese Academy of Sciences, Beijing 100049, China}
\affiliation{CAS Key Laboratory of High Precision Nuclear Spectroscopy, Institute of Modern Physics, Chinese Academy of Sciences, Lanzhou 730000, China}
\author{Xingbo Zhao}
\affiliation{Institute of Modern Physics, Chinese Academy of
Sciences, Lanzhou 730000, China}
\affiliation{School of Nuclear Science and Technology, University of Chinese Academy of Sciences, Beijing 100049, China}
\affiliation{CAS Key Laboratory of High Precision Nuclear Spectroscopy, Institute of Modern Physics, Chinese Academy of Sciences, Lanzhou 730000, China}
\author{James P. Vary}
\affiliation{Department of Physics and Astronomy, Iowa State University, Ames, IA 50011, USA}

\begin{abstract}
We employ the non-perturbative time-dependent basis function (tBF) approach to study the scattering of the deuteron on $^{208}$Pb below the Coulomb barrier. We obtain the bound and discretized scattering states of the projectile, which form the basis representation of the tBF approach, by diagonalizing a realistic Hamiltonian in a large harmonic oscillator basis. We find that the higher-order inelastic scattering effects are noticeable for sub barrier scatterings with the tBF method. We have successfully reproduced experimental sub Coulomb barrier elastic cross section ratios with the tBF approach by considering only the electric dipole ({\it E1}) component of the Coulomb interaction between the projectile and the target during scatterings. We find that the correction of the polarization potential to the Rutherford trajectory is dominant in reproducing the data at very low bombarding energies, whereas the role of internal transitions of the deuteron projectile induced by the {\it E1} interaction during the scattering becomes increasingly significant at higher bombarding energies.
\end{abstract}

 \maketitle

\section{Introduction}
The electric dipole ({\it E1}) polarizability, which is one of the fundamental properties of a nucleus, has been the subject of extensive experimental measurements and theoretical calculations.
The first experimental determination of the {\it E1} polarizability for the deuteron was achieved, by  measuring a quantity $R(E_d)$ (defined in terms of the ratios of elastic scattering cross sections at different scattering angles) of the deuteron scattering on $^{208}$Pb at energies well below the Coulomb barrier~\cite{Rodning:1982zz}. Comparing the experimental $R(E_d)$ and the results of the optical model analyses with deuteron incident energies $E_d=3-7$ MeV, Ref.~\cite{Rodning:1982zz} obtained an empirical value of {\it E1} polarizability of the deuteron $\alpha=0.7\pm0.05$ fm$^3$. An independent empirical value of the {\it E1} polarizability of the deuteron, $\alpha=0.61\pm0.04$ fm$^3$, has been extracted from deuteron photoabsorption data~\cite{Friar:1983zza}.  The uncertainties of these two empirical data are sufficiently large that the two results are nearly consistent with each other. Theoretical values of the {\it E1} polarizability of the deuteron, calculated with various nucleon-nucleon ({\it NN}) interactions, range from $0.6$ fm$^3$ to $0.65$ fm$^3$~\cite{Friar:1984zzb,Friar:1997gv}. Future improvements to the experimental dipole polarizability of the deuteron could be useful to provide additional constraints on the {\it NN} interaction.

The optical model has been used to analyze the experimental $R(E_d)$ employing various phenomenological optical potentials~\cite{Rodning:1982zz,Aoki:2000nim,Moro:1999fcl}. Although all of these calculations are able to reproduce the experimental data, their reaction dynamics differ from each other, raising questions about our understanding of the relevant reaction processes. For example, the optical potential contributes more than $48\%$ to the reduction of $R(E_d)$ at $E_d=7$ MeV in Ref.~\cite{Rodning:1982zz}, while Ref.~\cite{Moro:1999fcl} shows that the contribution of the optical potential is less than $33\%$.
%One of the reasons for such discrepancies is the uncertainty of the parameters in the optical potentials.
The optical potentials between the deuteron projectile and the $^{208}$Pb target adopted in these optical model calculations are obtained by fitting experimental data at energies above our range of interest, i.e., above $E_d=7$ MeV, ~\cite{Rodning:1982zz,Moro:1999fcl,Daehnick:1980zz} or by an extrapolation from high energies to low energies~\cite{Moro:1999fcl}. The significance of the optical potential in explaining the experimental $R(E_d)$ is therefore a matter of some disagreement in the literature.

The conventional optical model underlines a single channel analysis where the absorption due to internal transitions from the deuteron ground state to the scattering states, induced by the Coulomb interaction between the projectile and the target, are accommodated by its imaginary part. In Ref.~\cite{Aoki:2000nim} the continuum discretized coupled channel (CDCC) method takes specified transitions into account explicitly and also includes an imaginary part of the optical potential. The CDCC method also reproduces the experimental $R(E_d)$ successfully while the contribution of the explicitly included transitions of the deuteron projectile to $R(E_d)$ was not presented independent of the imaginary part of the optical potential.

In this work, we calculate the quantity $R(E_d)$ of the deuteron scattering on $^{208}$Pb using a non-perturbative time-dependent scattering approach, considering only the Coulomb interaction between the deuteron projectile and the $^{208}$Pb target. At the sub Coulomb barrier energies, we demonstrate that the induced {\it E1} transitions to and among the excited states of the deuteron lead to a successful description of the experimental $R(E_d)$ without the need for an optical potential.

In Refs.~\cite{Du:2018a,Du:2018b}, we proposed the theoretical framework for the time-dependent basis function (tBF) approach to nuclear scattering processes. The aim of the tBF method is to extend the {\it ab initio} nuclear structure approaches, e.g., the no-core shell model (NCSM)~\cite{Barrett:2013nh}, to microscopic nuclear reaction theory. The key idea for this extension is the construction of the basis representation for the scattering problem, in which the solutions of the {\it ab initio} nuclear structure approaches are encoded into the reaction theory. Within this basis representation, the equation of motion (EOM) of the scattering process is solved numerically as an initial value problem in a non-perturbative manner, where the quantal coherence is fully retained during the complicated reaction process. In addition to the reaction observables (e.g., the cross section), the detailed dynamics and various quantal phenomena (e.g., the entanglement between reaction fragments) for the complicated scattering processes can be investigated. We adopt the deuteron as the incident nucleus for our example case since it is the simplest nucleus which is numerically accessible employing realistic {\it NN} interactions. Future applications will use {\it ab initio} methods to describe other incident projectiles.

In this work, we improve the tBF method introduced in Refs.~\cite{Du:2018a,Du:2018b} to investigate the scattering of the deuteron projectile on the $^{208}$Pb target well below the Coulomb barrier of approximately $11$ MeV. The $^{208}$Pb target provides a static Coulomb field that acts on the deuteron projectile.  This external field causes the scattering of the center of mass (COM) of the projectile, a polarization of the projectile which influences the COM motion, and breakup through inelastic {\it E1} excitations of the projectile. Our advances to the tBF method in the present work are two-fold. First, we construct the basis representation of the tBF method with both the bound and scattering states of the deuteron obtained by diagonalizing a realistic Hamiltonian in a large three-dimensional (spherical) harmonic oscillator (HO) basis following the NCSM methodology. Second, we employ a realistic classical scattering trajectory for the COM of the projectile, which is determined by both the Coulomb potential and the polarization potential~\cite{Rodning:1982zz,Aoki:2000nim,Moro:1999fcl}.

%In Refs.~\cite{Du:2018a,Du:2018b}, we proposed the {\it ab initio} non-perturbative, time-dependent basis function (tBF) method to investigate the dynamics of the Coulomb excitation of the deuteron (trapped with an external harmonic oscillator potential) by an impinging heavy ion. The tBF approach retains the full quantal coherence and therefore can be used to study the detailed dynamics for complicated scattering processes. In this work, we investigate the scattering of a deuteron projectile on a $^{208}$Pb target below the Coulomb barrier (which is approximately $11$ MeV) with an improved tBF method. We solve the equation of motion of the tBF method in the basis representation formed by the bound and breakup channels of the deuteron obtained by diagonalizing a realistic Hamiltonian in a large three-dimensional (spherical) harmonic oscillator (HO) basis. We regard the $^{208}$Pb target as a classical source of a static external Coulomb field acting on the center of mass (COM) of the deuteron and a source of the electric dipole ({\it E1}) transitions within the deuteron system. This work is a pioneering development of the tBF method before we generalize this approach to treat more complicated problems, for example, the scattering of other light projectiles on heavy targets, where the basis representation of the tBF method will be formed by the eigenstates of the light projectiles solved with the {\it ab initio} NCSM~\cite{Barrett:2013nh}.
%

The present paper is organized as follows. In Sec. \ref{sec:theoryReview}, we present the theoretical framework of this paper. We present and discuss the results in Sec. \ref{sec:resultsAndDiscussion}. Finally, we give a summary of our conclusions in Sec. \ref{sec:conclusions}.

\section{Theoretical framework}
\label{sec:theoryReview}
In this paper we adopt the tBF approach to study the scattering of the deuteron projectile on the $^{208}$Pb target below the Coulomb barrier. Detailed descriptions of this approach can be found in Refs.~\cite{Du:2018a,Du:2018b}. Here we simply present a brief review for completeness. We also introduce an extension of the previous work to include a polarization potential acting on the deuteron.

\begin{figure}[tbh]
\begin{center}
\includegraphics[width=0.7\textwidth]{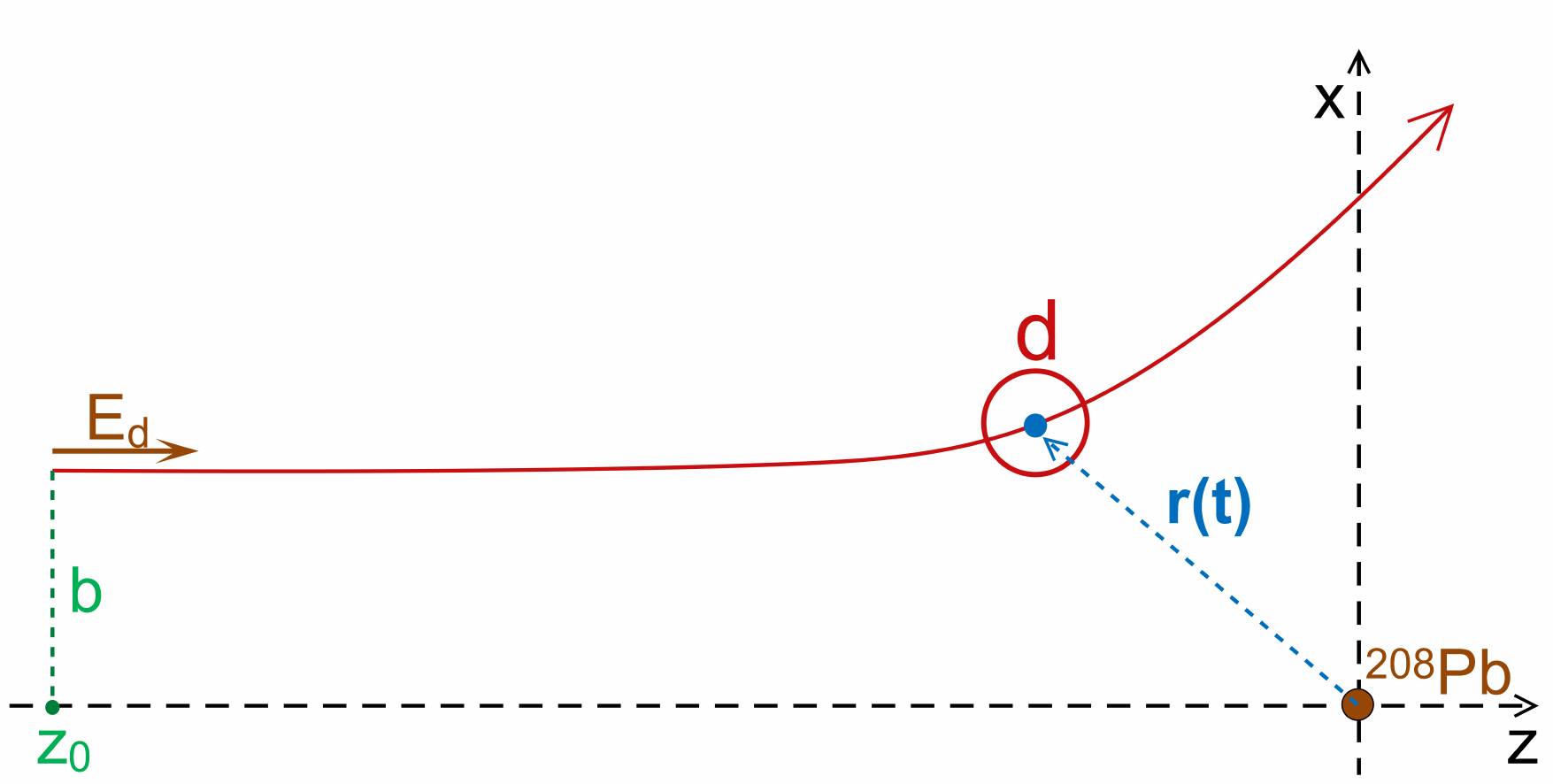}
\end{center}
\caption{(Color online) A sketch for the scattering of the deuteron projectile on the $^{208}$Pb target. See the text for the details.}\label{fig1}
\end{figure}

%\textcolor[rgb]{0.00,0.00,1.00}{\subsection{Setup of the scattering}}
The sketch of the scattering setup is presented in Fig.~\ref{fig1}. The scattering plane is taken to be the $xz$ plane. The bare $^{208}$Pb (with all electrons removed) target is fixed at the origin (we work in the lab frame so the equivalent assumption is that the target is infinitely massive). For simplicity, we take it to be a point-like nucleus in this work. The initial velocity of the deuteron projectile is parallel to the $z$ axis, with $E_d$ the corresponding bombarding energy. We treat the COM of the projectile as moving along a classical trajectory, which is determined by the interaction between the projectile and the target. $b$ denotes the impact parameter, which is determined according to Ref.~\cite{Baur:1977nso} for a given bombarding energy and scattering angle. The time-dependent vector ${\bf r}(t)$ denotes the position of the COM of the neutron-proton ({\it np}) system with respect to the origin during the scattering. In numerical simulations for the classical trajectory, the initial separation between the projectile and the target will be taken to be large but finite. We calculate the trajectory assuming that the deuteron projectile is initially located at $(x, z)=(b,z_0)$ at $t=0$ as shown in Fig.~\ref{fig1}. For the bombarding energies and scattering angles of the present work, we adopt $z_0 = -300$ MeV$^{-1}$ and discuss this choice below.

%\textcolor[rgb]{0.00,0.00,1.00}{\subsection{Theory of the tBF method}}
The full Hamiltonian of the {\it np} system moving in the time-dependent background field produced by $^{208}$Pb can be written as
\begin{eqnarray}
H_{\rm full}(t) &=& H_0 + V_{\rm int}(t) \label{eq:FullH}  ,
\end{eqnarray}
where $V_{\rm int}(t)$ denotes the time-dependent interaction between the projectile and the target. $H_0$ denotes the ``free'' Hamiltonian for the intrinsic motion of the {\it np} system:
\begin{eqnarray}
H_0 &=& T_{\rm rel} + V_{\rm NN}   \label{eq:H0}  ,
\end{eqnarray}
with $T_{\rm rel}$ and $V_{\rm NN}$ being the relative kinetic energy and the {\it NN} interaction, respectively.
%Note that we remove the external harmonic oscillator trap applied to regulate the scattering states in our previous work \cite{Du:2018a,Du:2018b}.
%In Ref.~\cite{Du:2018b}, an external harmonic oscillator trap is introduced to the Hamiltonian for the intrinsic motion of the {\it np} system to regularize the scattering states. In this paper we remove the external harmonic oscillator to model the scattering states of the {\it np} system.

The eigenstates, both the bound state and the scattering states, of the projectile can be solved from the eigenequation,
\begin{eqnarray}
H_{0} |\beta_j \rangle &=& E_{j} \ |\beta_j \rangle \label{eq:EFunc}  ,
\end{eqnarray}
where $E_j$ and $|\beta_j \rangle$ represent the eigenvalue and the corresponding eigenvector, respectively. The subscript $j$ is an index for the bound and scattering states.
In practice, we adopt the HO representation to solve Eq.~(\ref{eq:EFunc}). The parameters of the HO basis include the basis strength $\omega$ and the basis truncation parameter $N_{\rm max}$ (defined as the maximum of twice the radial quantum number plus the orbital angular momentum)~\cite{Barrett:2013nh,Vary:2018jxg}.
%In the following, we denote the strength of the HO basis by $\omega$. In practice, it is advantageous to define the truncation to the space of the HO basis with $N_{max}$~\cite{Vary:2018jxg,Barrett:2013nh}.
Once the basis size is sufficiently large (scaled by $N_{\rm max}$), the lowest lying state coincides with the deuteron bound state, while all the other excited states are regarded as a discretized approximation of the continuum~\cite{Dinur:2014kha,Hernandez:2014}.

The EOM of the projectile during the scattering, in the interaction picture, can be written as
\begin{eqnarray}
i \frac{\partial}{\partial t}|\psi; t \rangle _I &=&  e^{i {H_{0}t}}\ V_{\rm int}(t)\ e^{-i {H_{0}}t}\ |\psi; t \rangle _I \ \equiv \  V_I(t)\ |\psi; t \rangle _I \ , \label{eq:EOMequation}
\end{eqnarray}
where $V_I(t)$ denotes the time-dependent interaction between the projectile and the target in the interaction picture. The subscript ``I'' specifies the interaction picture. By virtue of using $H_0$ to generate the time evolution, we are including the interactions of the {\it np} system in the intermediate and final states involved in the scattering. Note that we adopt the natural units and set $\hbar = c = 1$ throughout this paper. For the tBF method, we solve the state vector $| \psi ; t \rangle _I$ via the non-perturbative multistep differencing scheme up to the second-order (MSD2)~\cite{Iitaka:1994} in the basis representation formed by the set of state vectors $\{|\beta_j\rangle\}$ in Eq.~(\ref{eq:EFunc}).

 In the present paper, we consider only the {\it E1} component of the Coulomb interaction between the projectile and the target since that is known to be the dominant deuteron excitation mode for sub barrier scatterings~\cite{Moro:1999fcl}. During the scattering, the time-dependent interaction $V_I(t)$ [Eq.~(\ref{eq:EOMequation})] induces the {\it E1} transitions of the projectile~\cite{Du:2018a,Du:2018b}. Hence, the inelastic effects in this paper stem entirely from {\it E1} transitions. We do not consider the excitation of $^{208}$Pb since its spin-parity ($3^-$) requires a higher multipole ({\it E3}) transition and its excitation energy ($2.6$ MeV) is also substantial~\cite{Aoki:2000nim}. Hence, as previously noted~\cite{Rodning:1982zz}, excitation of $^{208}$Pb is expected to be two orders of magnitude smaller than the Coulomb dissociation of the deuteron at these sub Coulomb barrier energies. The influence of vacuum polarization, atomic screening and relativistic corrections on the quantity $R(E_d)$ are also found to be small~\cite{Rodning:1982zz,Moro:1999fcl} and therefore these effects are not taken into account in the present calculation.

 The contribution of the neutron stripping reactions ($d$,$p$) to $R(E_d)$  below $E_d=7$ MeV has not been determined in experiment. In optical model calculations of elastic scattering, the contribution of neutron stripping reactions would be accounted for, along with breakup to {\it np} final states, by the imaginary component of the optical potential. Since we treat all inelastic processes together (inclusive inelastic scattering), we are, in principle, including processes that feed the ($d$,$p$) process together with processes that feed the {\it np} pair in the final state continuum. That is, the deuteron inelastic processes we include could feed neutron transfer to $^{208}$Pb which would be quantum mechanically indistinguishable from a direct neutron transfer reaction to the same final state. We will investigate these coupled reaction channels in a future extension to the tBF method. In the meantime, experimental measurements with high resolution and comparisons with our theoretical calculations for inclusive inelastic scattering, that we currently investigate, promise to provide constraints to theoretical approaches in the future.

 We examined the effect of the magnetic dipole ({\it M1}) transitions of the {\it np} system induced by the time-dependent electromagnetic interaction between the projectile and the target with the tBF method. We found that the effects of the {\it M1} transitions were negligibly small compared to the effects of the {\it E1} transitions below the Coulomb barrier and would not affect the conclusions of this paper so we omit the {\it M1} transitions at the present time.

%\textcolor[rgb]{0.00,0.00,1.00}{\subsection{Correction of E1 polarization potential to Rutherford trajectories}}

 For scattering well below the Coulomb barrier, the Rutherford trajectory is thought to be a good first-order approximation. The Rutherford trajectory is determined by the following Coulomb potential
 \begin{eqnarray}
V_{\rm c}=\frac{Ze^2}{r(t)},
\label{eq:Vpot_Coloumb}
\end{eqnarray}
 where $Z$ represents the charge number of the target ($Z=82$ in this paper). However, the Coulomb field produced by the target also polarizes the projectile and this leads to a correction to the Coulomb potential. This effect can be taken into account by a polarization potential $V_{\rm pol}$.

We follow Refs.~\cite{Rodning:1982zz,Moro:1999fcl,Baur:1977nso} and employ a polarization potential obtained from second-order perturbation theory, which is written as
 \begin{eqnarray}
 V_{\rm pol}=-\frac{1}{2}\alpha\frac{Z^2e^2}{r^4(t)}.
 \label{eq:VSOP}
 \end{eqnarray}
 $\alpha$ is the {\it E1} polarizability of the deuteron which is defined as~\cite{Alder:1956im}
 \begin{eqnarray}
\alpha &=& \frac{8\pi}{9}\sum_{n\ne0}\frac{B(E1;0\rightarrow n)}{(E_n-E_0)},
\label{eq:alpha}
\end{eqnarray}
 where the indexes $0$ and $n$ denote the ground state and the {\it E1} excited states of the deuteron, respectively. $B(E1;0\rightarrow n)$ represents the electric dipole strength for the coupling between the deuteron ground state $|\beta_0\rangle$ and the {\it E1} excited state $|\beta_n\rangle$ which is calculated as follows~\cite{Alder:1956im}
 \begin{equation}
B(E1;0\rightarrow n)
=\sum_{M_n,\mu}\left|\left\langle\beta_0,M_0|\mathcal{M}(E1,\mu)|\beta_n,M_n\right\rangle\right|^2,
\label{eq:BE1}
\end{equation}
where $\mathcal{M}(E1,\mu)$ denotes the {\it E1} operator. $M_0$ and $M_n$ represent the orientations of the ground state $|\beta_0\rangle$ and the {\it E1} excited state $|\beta_n\rangle$, respectively.
We then solve for the trajectory of the COM of the projectile with the combined potential,
\begin{eqnarray}
V_{\rm pot} &=& V_{\rm c}+V_{\rm pol}.
\label{eq:Vpot}
\end{eqnarray}

%\textcolor[rgb]{0.00,0.00,1.00}{\subsection{cross section}}
As in Refs.~\cite{Esbensen:2008zp,Aleixo:1989cb,Alder:1975}, we evaluate the differential cross section of the elastic scattering as
\begin{equation}
\left(\frac{d\sigma}{d\Omega}\right)_{\rm el} = P_{\rm el}\left(\frac{d\sigma}{d\Omega}\right)_{\rm class}, \label{eq:CSection}
\end{equation}
where $P_{\rm el}$ denotes the elastic scattering probability and is obtained by summing over the probabilities in the three orientations of the deuteron ground state after the time evolution with the tBF method. The classical differential cross section $\left(\frac{d\sigma}{d\Omega}\right)_{\rm class}$ is calculated using a trajectory defined by the adopted potential acting on the COM of the deuteron (either $V_{\rm c}$ or $V_{\rm pot}$) and is represented by
\begin{eqnarray}
\left(\frac{d\sigma}{d\Omega}\right)_{\rm class} &=&  \frac{b}{\sin\theta}\left|\frac{db}{d\theta}\right|,
\end{eqnarray}
where $b$ and $\theta$ denote the impact parameter and the scattering angle, respectively. For reference, in the case where $V_{\rm c}$ alone is used, the Rutherford cross section would emerge since $b=\frac{Ze^2}{2E_d}\cot \left(\frac{\theta}{2}\right)$.
In light of available experimental data~\cite{Rodning:1982zz} we calculate the following quantity $R(E_d)$
\begin{eqnarray}
R(E_d)=\frac{\sigma(3\ \rm{MeV}, \theta_1)}{\sigma(3\ \rm{MeV}, \theta_2)}\frac{\sigma(E_d, \theta_2)}{\sigma(E_d, \theta_1)},
\label{eq:RE}
\end{eqnarray}
where $\sigma(E_d, \theta)=2\pi\left(\frac{d\sigma}{d\Omega}\right)_{\rm el}$ denotes the differential cross section of the elastically scattered deuterons at angle $\theta$ with the bombarding energy $E_d$.

\section{Results and Discussions}
\label{sec:resultsAndDiscussion}

 In this work, we solve Eq.~(\ref{eq:EFunc}) with the {\it NN} interaction constructed from the chiral effective field theory to obtain the bound and breakup states of the deuteron projectile. In particular, we employ an {\it NN} interaction of the Low Energy Nuclear Physics International Collaboration (LENPIC)~\cite{Epelbaum:2014sza,Epelbaum:2014efa,Maris:2016wrd,Binder:2016,Binder:2018} up to N$^4$LO (which we refer to as LENPIC-N$^4$LO). The LENPIC interactions employ a semilocal coordinate-space regulator and we adopt the interaction with the regulator of $1.0$ fm~\cite{Binder:2016,Binder:2018}. In the tBF method, the deuteron spectrum, {\it E1} transition matrix elements and {\it E1} polarizability depend on the adopted {\it NN} interaction. Hence, the sensitivity of scattering observables to the {\it NN} interaction is likely to depend on the incident energy and the scattering angles. We will test other {\it NN} interactions in future applications in order to investigate tBF scattering conditions that could constrain the off-shell properties of the realistic {\it NN} interaction.

%\textcolor[rgb]{0.00,0.00,1.00}{\subsection{Choice of scattering channels}}
  We set the initial state of the projectile to be in its ground state ($^3S_1-{^3D_1}$ channel). The polarization will be defined for each of the specific applications below. Since {\it E1} transitions respect the conservation of the total spin $S$ of the {\it np} system, we take only channels with $S=1$ into account. We restrict the total angular momentum $J$ to be $J\le 2$ though higher angular momentum states could, in principle, be populated through higher-order transitions. We introduce a quantity $E_{\rm cut}$ to represent the upper energy limit of the retained scattering states of the {\it np} system. We discuss below our choice $E_{\rm cut}=14$ MeV and its adequacy. To be specific, we adopt the eigenstates of the {\it np} system with eigenenergies below $E_{\rm cut}$ in $^3S_1-{^3D_1}$, $^3P_0$, $^3P_1$, $^3D_2$ and $^3P_2-{^3F_2}$ channels to form the basis representation of the tBF approach in this work.

In order to simplify numerical implementation of the tBF method and retain sufficient numerical precision simultaneously, we will employ a set of chosen computational elements ($z_0$, $r_c$ and $\delta t$ detailed below), where we adopt the same initial polarization of the deuteron, {\it NN} interaction, truncation parameters and polarization potential as in Fig. \ref{fig4}.

In this work, we obtain all the trajectories with $z_0=-300$ MeV$^{-1}$ ($\approx-59200$ fm). We deem this to be sufficient for the scattering observable $R(E_d)$ since the probability of the deuteron remaining in the ground state, after time evolution, at $E_d=7$ MeV and $\theta=150^\circ$, is changed by only around $10^{-5}$ with $z_0$ extended from $-300$ MeV$^{-1}$ to $-400$ MeV$^{-1}$ ($\approx-78900$ fm). For simplicity and to be conservative, we retain this initial value of $z_0$ for lower bombarding energies.
%We take the time step $\delta t=0.006$ MeV$^{-1}$ in these tBF calculations and initiate the internal transitions in the deuteron projectile at $t=0$ when the COM of the deuteron is located at $(x,z)=(b,z_0)$ as shown in Fig. \ref{fig1}.

To improve the computational efficiency, we will initiate the internal transitions from an intermediate moment $t_0$, after which the separation of the projectile and the target is reduced to be less than $r_c$ which we determine in the following manner. We found that the internal transitions in the deuteron projectile are negligible for sufficiently distant separation of the projectile and the target. Through numerical studies, we found that $r_c=5$ MeV$^{-1}$ ($\approx 987$ fm) is a reasonable choice since the probability of the deuteron remaining in the ground state, after scattering, at $E_d=7$ MeV and $\theta=150^\circ$, is converged up to the fifth significant figure comparing with the result calculated with $r_c=10$ MeV$^{-1}$ ($\approx 1973$ fm). Therefore, we adopt $r_c = 5$ MeV$^{-1}$ for our applications here.

The probabilities of the deuteron remaining in the ground state, after scattering, at $E_d=7$ MeV and $\theta=150^\circ$, using $\delta t=0.0001$ MeV$^{-1}$ and $\delta t=0.006$ MeV$^{-1}$, also agree within five significant figures. Therefore we adopt $\delta t=0.0001$ MeV$^{-1}$ in the following applications.

%\textcolor[rgb]{0.00,0.00,1.00}{\subsection{E1 polorizablility of the projectile}}
\begin{figure}[tbh]
\begin{center}
\includegraphics[width=0.8\textwidth]{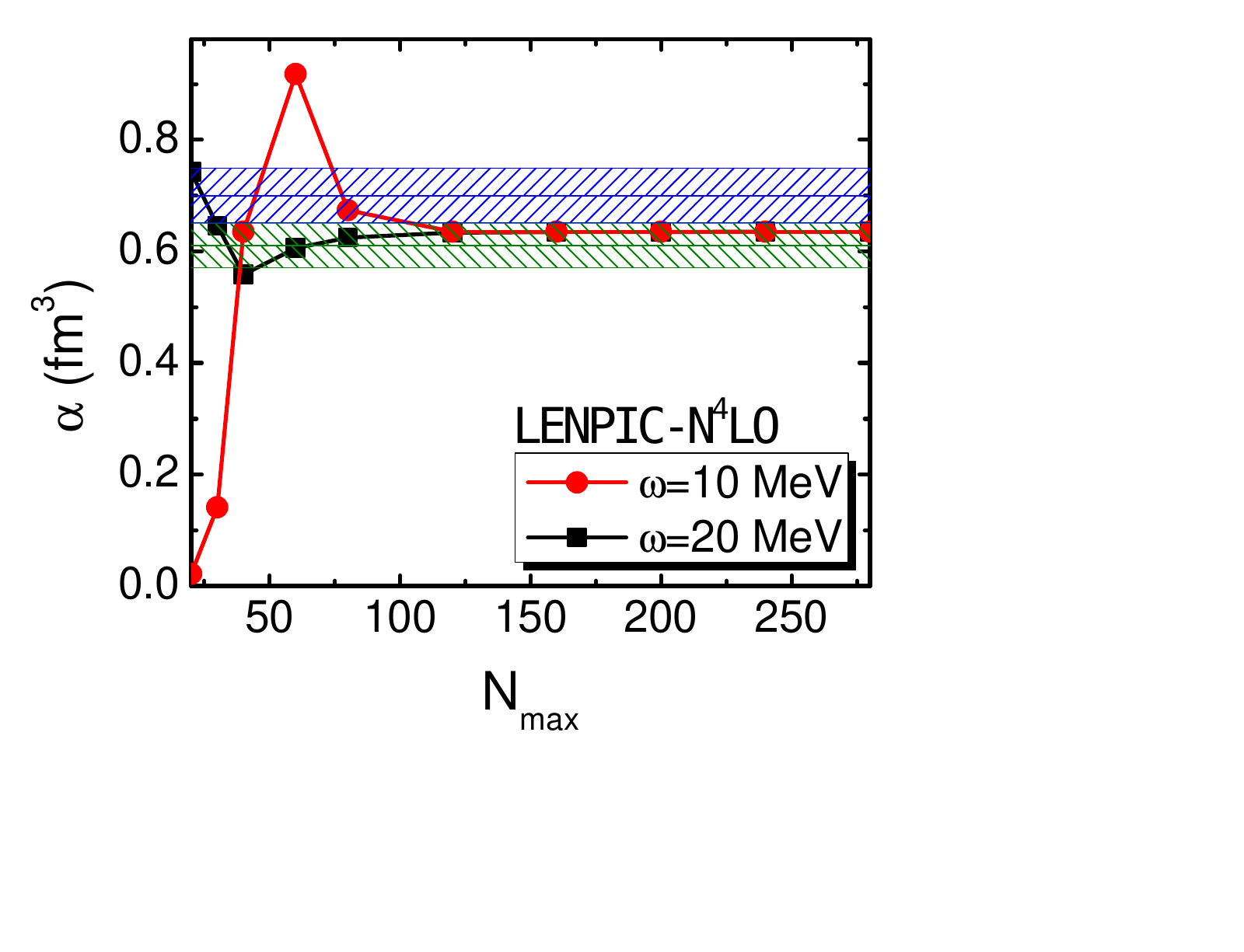}
\end{center}
\caption{(Color online) The {\it E1} polarizability of the deuteron, $\alpha$, which is calculated with the LENPIC-N$^4$LO interaction as a function of the truncation parameter $N_{\rm max}$. Results are shown for two different strengths of the HO basis, i.e., $\omega=10$ MeV (red solid circles) and $\omega=20$ MeV (black solid squares). Two experimental results and their uncertainty bands from Ref.~\cite{Rodning:1982zz} (blue region) and Ref.~\cite{Friar:1983zza} (olive region) are presented for comparison.}\label{fig_1}
\end{figure}
In Fig. \ref{fig_1}, we present the {\it E1} polarizability of the deuteron [see Eq.~(\ref{eq:alpha})] as a function of the truncation parameter $N_{\rm max}$ which is calculated with the LENPIC-N$^4$LO {\it NN} interaction for two basis strengths ($\omega=10$ and 20 MeV) of the HO basis. We also present two sets of results extracted from experiments along with their quoted uncertainties~\cite{Rodning:1982zz,Friar:1983zza} for comparison. We find from Fig. \ref{fig_1} that the {\it E1} polarizability of the deuteron predicted by the LENPIC-N$^4$LO interaction reaches a convergent value at sufficiently large $N_{\rm max}$ and that the {\it E1} polarizability is independent of $\omega$. The converged value $\alpha=0.635$ fm$^3$, which will be employed in the following tBF calculations, is consistent with the two results extracted from experimental data~\cite{Rodning:1982zz,Friar:1983zza}. The {\it E1} polarizability of the deuteron based on the LENPIC-N$^4$LO interaction is also close to the results predicted by other realistic {\it NN} interactions~\cite{Friar:1984zzb,Friar:1997gv}.

%\textcolor[rgb]{0.00,0.00,1.00}{\subsection{Demonstration: Level population according to the Rutherford trajectory}}
\begin{figure}[tbh]
\begin{center}
\includegraphics[width=1.\textwidth]{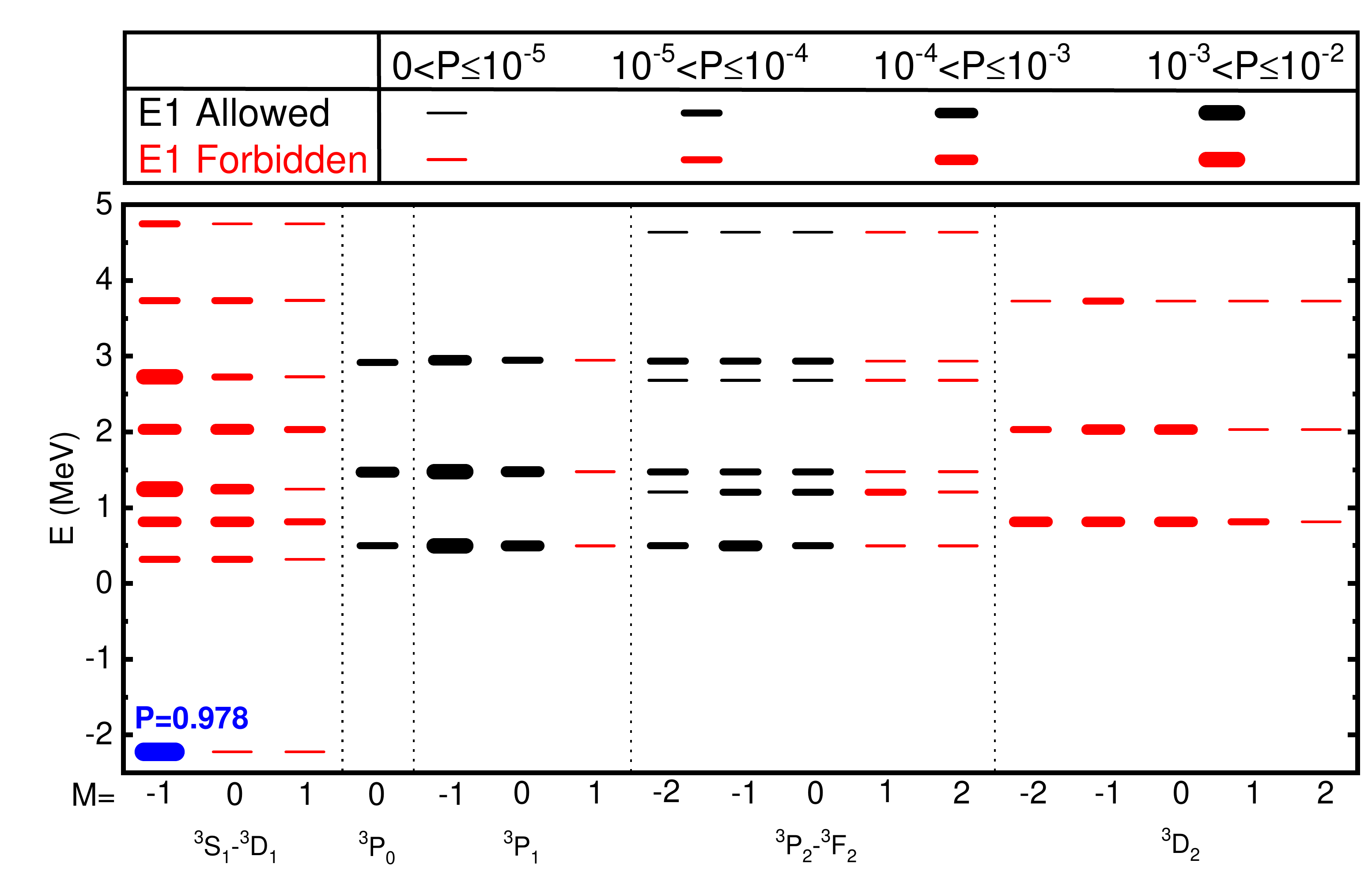}
\end{center}
\caption{(Color online) Energy levels of the ground state and scattering states in five channels ($^3S_1-{^3D_1}$, $^3P_0$, $^3P_1$, $^3D_2$ and $^3P_2-{^3F_2}$) of the {\it np} system predicted by the LENPIC-N$^4$LO interaction in HO basis with $\omega=20$ MeV and $N_{\rm max}=200$. After the scattering of d+$^{208}$Pb at $E_d=7$ MeV and $\theta=150^\circ$, the occupation probability of each state is calculated by the tBF method and denoted by its thickness as indicated in the legend. The {\it E1} allowed and forbidden states are distinguished by the black and red levels, respectively. The initial state (blue level with occupation probability $P=0.978$ after scattering) is taken to be ($^3S_1-{^3D_1}$, $M=-1$).}\label{fig3}
\end{figure}
In Fig. \ref{fig3}, we present the spectrum of the {\it np} system calculated by the LENPIC-N$^4$LO interaction in the HO basis with $\omega=20$ MeV. We take the truncation parameter of the HO basis to be $N_{\rm max}=200$. With the restriction of $E_{\rm cut}=14$ MeV, we obtain $165$ bound and discretized scattering states (counting each, possibly degenerate, state), in total, by solving Eq.~(\ref{eq:EFunc}). We investigate the scattering of the deuteron on $^{208}$Pb at $E_d=7$ MeV and $\theta=150^\circ$ by the tBF method employing the basis representation formed by these $165$ states. We note in passing that $165$ states are far from our computational limits but were found sufficient for our purposes in the present work. For the initial state we take a polarized deuteron in its ground state  ($^3S_1-{^3D_1}$, $M=-1$) which is the blue level in Fig. \ref{fig3}. In the calculation, we adopt the polarization potential described by Eq.~(\ref{eq:VSOP}). During the scattering, a network of {\it E1} transitions is activated by the external field with coherent transitions taking place among all $165$ states. That is, wherever a possible {\it E1} transition can occur, it participates in a multitude of excitation pathways in the network. Consequently, the population of each state, the absolute square of its amplitude,  changes with time and reaches the converged value after the {\it E1} interaction between the projectile and the target is negligibly small. The resulting populations of the continuum states shown in Fig. \ref{fig3} represent the coherent inelastic effects introduced in the tBF method. These final populations are represented by the thickness of the lines as defined in the legend.

After scattering, the sum of the populations of the kinematically forbidden states, i.e., those with excitation energy above $7$ MeV, is on the order of $10^{-5}$ which is negligibly small and is taken as the numerical uncertainty for each state's population. We have therefore presented the populations of states with excitation energy below $7$ MeV ($81$ states in total) in Fig. \ref{fig3}. We signify states allowed and forbidden by {\it E1} in first-order perturbation theory from the initial state by the black and red levels in Fig. \ref{fig3}, respectively. For simplicity we will refer to these states as either ``{\it E1} allowed" or ``{\it E1} forbidden" accordingly.

After the time evolution, we observe populations in all the {\it E1} forbidden states in Fig. \ref{fig3}. Around $45$ percent of the {\it E1} forbidden states are populated significantly above the level of numerical uncertainty. We find that the populations in several {\it E1} forbidden states are comparable to those in the {\it E1} allowed states, which indicates the significance of the higher-order inelastic scattering effects that emerge in these tBF calculations.

The total probability for populating breakup states presented in Fig. \ref{fig3} is around $0.022$. For contrast, we also perform a similar calculation but retain only the transitions between the initial state and the {\it E1} allowed states. We find that the resulting probability for populating breakup states, is around $0.00015$. Therefore, the contribution of the higher-order transitions to the inelastic effect is more than two orders of magnitude larger than that of the {\it E1} allowed transitions alone. We note that a similar dominance by higher-order effects was observed for the total reaction cross section in Ref.~\cite{Aoki:2000nim}.

%\textcolor[rgb]{0.00,0.00,1.00}{\subsection{Scattering observables: $R(E_d)$}}
\begin{figure}[tbh]
\begin{center}
\includegraphics[width=1.0\textwidth]{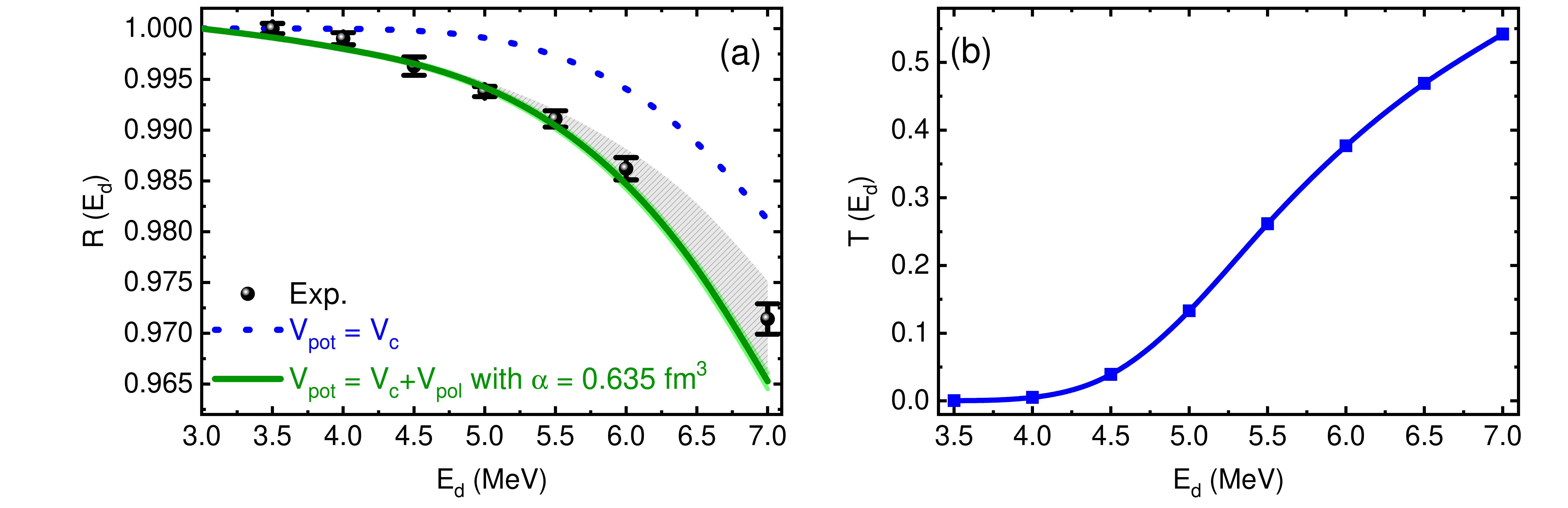}
\end{center}
\caption{(Color online) Elastic cross section ratios $R(E_d)$ with $(\theta_1,\theta_2)=(60^\circ,150^\circ)$ [panel (a)] and the quantity $T(E_d)$ measuring the contribution of the internal transitions to $1-R(E_d)$ [panel (b)], as functions of the bombarding energy $E_d$. The dark green solid curve and the blue dotted curve represent the tBF results with and without the correction of the polarization potential to the Rutherford trajectory, respectively. The light green band represents the uncertainty induced by a $5\%$ change of the {\it E1} polarizability of the deuteron. The light grey band denotes the uncertainty induced by energy loss during the scattering. The experimental data~\cite{Rodning:1982zz} of $R(E_d)$ (black solid dots with error bars in panel (a)) are also shown for comparison. See in the text for details.
%$R(E_d)$ is calculated by the tBF method employing different trajectories. The blue dotted curve is obtained with the Rutherford trajectory and the red solid curve is calculated by the modified Rutherford trajectory (with the correction of the $E1$ polarization potential). The experimental data (black solid dots with error bars) are also shown for comparison.
}\label{fig4}
\end{figure}

In the left panel of Fig. \ref{fig4} [panel (a)], we display the quantity $R(E_d)$ with $(\theta_1,\theta_2)=(60^\circ,150^\circ)$ [see Eq.~(\ref{eq:RE})] for the scattering of the deuteron on $^{208}$Pb at $E_d=3-7$ MeV calculated by the tBF method. For each $R(E_d)$ we calculate four cross sections, $\sigma(3\ \rm{MeV}, \theta_1=60^\circ)$, $\sigma(3\ \rm{MeV}, \theta_2=150^\circ)$, $\sigma(E_d, \theta_1=60^\circ)$ and $\sigma(E_d, \theta_2=150^\circ)$. We take the evenly weighted coherent sum of the magnetic substates of the deuteron ground state as the initial state. We also present the experimental data in Fig. \ref{fig4} for comparison~\cite{Rodning:1982zz}. We take the same LENPIC-N$^4$LO {\it NN} interaction and truncation parameters (i.e., $N_{\rm max}=200$ and $E_{\rm cut}=14$ MeV) as in Fig. \ref{fig3}.  We have checked the convergence of the quantity $R(E_d)$ with respect to $N_{\rm max}$ and $E_{\rm cut}$ by increasing each over a $20$ percent range (i.e., $N_{\rm max}=240$ and $E_{\rm cut}=16.8$ MeV). We find that the most significant change of $R(E_d)$ is on the order of $10^{-4}$ (for $E_d=7$ MeV). Therefore the results in Fig. \ref{fig4} are numerically accurate within the resolution of the graph with the present choice of $N_{\rm max}$ and $E_{\rm cut}$. Such convergence tests demonstrate that the tBF method presents a robust description of the underlying physics investigated in this work. However, we may anticipate convergence challenges using the HO basis as we move to treat heavier projectiles so we will consider alternative basis spaces, such as natural orbital basis~\cite{Constantinou:2016urz}, which should provide superior convergence properties.

We obtain the dark green solid curve in panel (a) of Fig. \ref{fig4} by adopting  trajectories which are determined by the Coulomb potential supplemented with the polarization potential, i.e., $V_{\rm pot}=V_{\rm c}+V_{\rm pol}$. We take the {\it E1} polarizability in the polarization potential to be $\alpha=0.635$ fm$^3$ (the converged value presented in Fig. \ref{fig_1} for this {\it NN} potential). The light green error band is evaluated by introducing a $5$ percent change to $\alpha$. That is, the upper (lower) boundary is obtained with $\alpha=0.603$ fm$^3$ ($\alpha=0.667$ fm$^3$). The light grey band represents the uncertainty induced by the correction of energy loss to the classical trajectory of the deuteron projectile, which will be explained in detail in the following. In panel (a) of Fig. \ref{fig4}, we find that our tBF results reproduce the experimental data for $E_d=3-6$ MeV while falling below experiment at $7$ MeV.

For comparison, we also plot in panel (a) of Fig. \ref{fig4} the quantity $R(E_d)$ predicted by the tBF method with the Rutherford trajectories which are not corrected by the effects of the polarization potential (blue dotted line). We find this calculation is not able to describe the experimental data. This suggests the correction to the Rutherford trajectory arising from the polarization potential is crucial for reproducing the experimental data.

For the classical Rutherford scattering, we have $\left(\frac{d\sigma}{d\Omega}\right)_{\rm el}=\left(\frac{d\sigma}{d\Omega}\right)_{\rm R}$ since $P_{\rm el}=1$ and $\left(\frac{d\sigma}{d\Omega}\right)_{\rm class}=\left(\frac{d\sigma}{d\Omega}\right)_{\rm R}$ [Eq.~(\ref{eq:CSection})] where $\left(\frac{d\sigma}{d\Omega}\right)_{\rm R}$ represents the Rutherford differential cross section. Based on the Rutherford scattering formulae, it is easy to see that $\frac{\sigma(E_d, \theta_1)}{\sigma(E_d, \theta_2)}$ in Eq.~(\ref{eq:RE}) is independent of $E_d$ and hence $R(E_d)=1$ for the Rutherford scattering. Therefore the deviation of the quantity $R(E_d)$ from unity, i.e., $1-R(E_d)$, indicates the deviation of a scattering from the classical Rutherford scattering.

In the following analyses we denote the quantity on the dark green solid line in Fig. \ref{fig4} (a) as $R_a(E_d)$ and the quantity on the blue dotted line in Fig. \ref{fig4} (a) by $R_b(E_d)$ for convenience. For the tBF results with the correction of the polarization potential to the Rutherford trajectories (dark green solid line), $1-R_a(E_d)$ is induced by the following effects:
 \begin{enumerate}
  \item internal transitions of the projectile induced by the {\it E1} interaction between the projectile and the target during the scattering which lead to $P_{\rm el}<1$,
  \item the correction of the polarization potential to classical Rutherford trajectories which gives rise to $\left(\frac{d\sigma}{d\Omega}\right)_{\rm class}<\left(\frac{d\sigma}{d\Omega}\right)_{\rm R}$.
\end{enumerate}
However, for the tBF approach without the correction of the polarization potential (blue dotted line), $1-R_b(E_d)$ is purely induced by the internal {\it E1} transitions in the projectile. We also find that $P_{\rm el}$ in both cases with (dark green solid curve) and without (blue dotted line) the corrections of the polarization potential are nearly the same. This signifies that the effects of the internal transitions are very similar and can be measured by $1-R_b(E_d)$ in both cases. Therefore we can examine the significance of the internal transitions (out of the above two effects) in generating $1-R_a(E_d)$ with the quantity $T(E_d)=\frac{1-R_b(E_d)}{1-R_a(E_d)}$ which is presented in panel (b) of Fig. \ref{fig4}. The larger the quantity $T(E_d)$ is, the more the internal transitions contribute relative to the polarization potential. We find from panel (b) that the effect of the internal {\it E1} transitions of the projectile on $1-R_a(E_d)$ is negligibly small at very low bombarding energies compared to the effect of the polarization potential. We also find that the contribution of the internal transitions of the projectile to $1-R_a(E_d)$ increases with the bombarding energy and becomes dominant at $E_d=7$ MeV, although both effects mentioned above are enhanced.

We can distinguish our results from those of previous analyses. Ref.~\cite{Rodning:1982zz} shows that the polarization potential plays the dominant role in explaining the experimental $R(E_d)$ below $6$ MeV, which is later confirmed by Ref.~\cite{Moro:1999fcl}. Our result is consistent with this conclusion of these two papers. However, the contributions of the optical potential and the polarization potential are found to be comparable in Ref.~\cite{Rodning:1982zz} for reproducing the experimental $R(E_d)$ at $E_d=7$ MeV, while the role of the optical potential is much less significant than that of the polarization potential in Ref.~\cite{Moro:1999fcl}. We are closer to Ref.~\cite{Moro:1999fcl} since we have not included an optical potential and we find that the non-perturbative internal {\it E1} transitions of the projectile play the dominant role in approximately reproducing the experimental data.

The optical model calculations in Refs.~\cite{Rodning:1982zz,Moro:1999fcl} do not take internal transitions of the deuteron projectile into account explicitly. In the tBF calculations, we include the internal transitions of the deuteron projectile in a non-perturbative manner. Nevertheless, we find, in agreement with Refs.~\cite{Rodning:1982zz,Moro:1999fcl}, that the polarization potential produces the largest effect at energies below $6.5$ MeV as shown in Fig. \ref{fig4} (b). However, due to the rapid increase in the higher-order {\it E1} transitions, the effect of the polarization potential in the tBF method is smaller than that found in Refs.~\cite{Rodning:1982zz,Moro:1999fcl} at $E_d=7$ MeV.

Note that we adopt the same polarization potential as in Ref.~\cite{Rodning:1982zz}, which is unregulated and calculated with Eq.~(\ref{eq:VSOP}). On the other hand, the unregulated polarization potential and a regulated dynamical polarization potential are employed in Ref.~\cite{Moro:1999fcl}. The regulator in the dynamical polarization potential in Ref.~\cite{Moro:1999fcl} corresponds to a range that is half of the closest approach for the head-on collision. The minimum closest approach for this work is $17.2$ fm (for $E_d=7$ MeV and $\theta=150^\circ$) where the unregulated polarization potential is about $-0.0351$ MeV and the correction of the regulation to the polarization potential is less than $0.0007$ MeV, which is negligibly small. We can therefore infer that the effect of the regulation in the polarization potential on our results would be insignificant, which is consistent with Ref.~\cite{Moro:1999fcl}. The contribution of the regulation in the polarization potential would likely be more significant when investigating Coulomb dissociations at higher incident energies.

\begin{center}
\begin{table}[h]
\renewcommand\arraystretch{1.5}
\caption{\label{tab:table1} $\chi^2$/data calculated by the tBF method. The second column corresponds to the result without the correction of the polarization potential. The third column is obtained with the tBF method using the polarization potentials with $\alpha=0.635$ fm$^3$.}
\setlength{\tabcolsep}{7mm}{
\begin{tabular}{c c c}
  \hline\hline
   $V_{\rm pol}$ & $0$ & $\alpha=0.635$ fm$^3$ \\
  $\chi^2$/data & 43.0 & 3.6 \\
  \hline  \hline
\end{tabular}}
\end{table}
\end{center}

To quantify the significance of the polarization potential in reproducing the experimental $R(E_d)$, we present in Table~\ref{tab:table1} the quantity $\chi^2$/data with and without the polarization potential. We define $\chi^2$/data as follows
\begin{eqnarray}
\chi^2/{\rm data}=\frac{1}{7}\sum_{i=1}^{7}\left[\frac{R^{\rm tBF}_i(E_d)-R^{\rm Exp}_i(E_d)}{\Delta R^{\rm Exp}_i(E_d)}\right]^2,
\label{eq:chisquare}
\end{eqnarray}
where $R^{\rm Exp}_i(E_d)$ denotes the central value of the experimental $R(E_d)$ in Fig. \ref{fig4} (a). $R^{\rm tBF}_i(E_d)$ represents the tBF results. $\Delta R^{\rm Exp}_i(E_d)$ corresponds to the experimental error in Fig. \ref{fig4} (a). The subscript $i$ runs over the seven points in Fig. \ref{fig4} (a). Comparing the results with and without the polarization potential in Table~\ref{tab:table1}, we find that the inclusion of the polarization potential in the tBF method significantly improves the description of the experimental data. To be more explicit, we find that the polarization potential provides a factor of $\sim12$ improvement in $\chi^2$/data.

\begin{figure}[tbh]
\begin{center}
\includegraphics[width=0.7\textwidth]{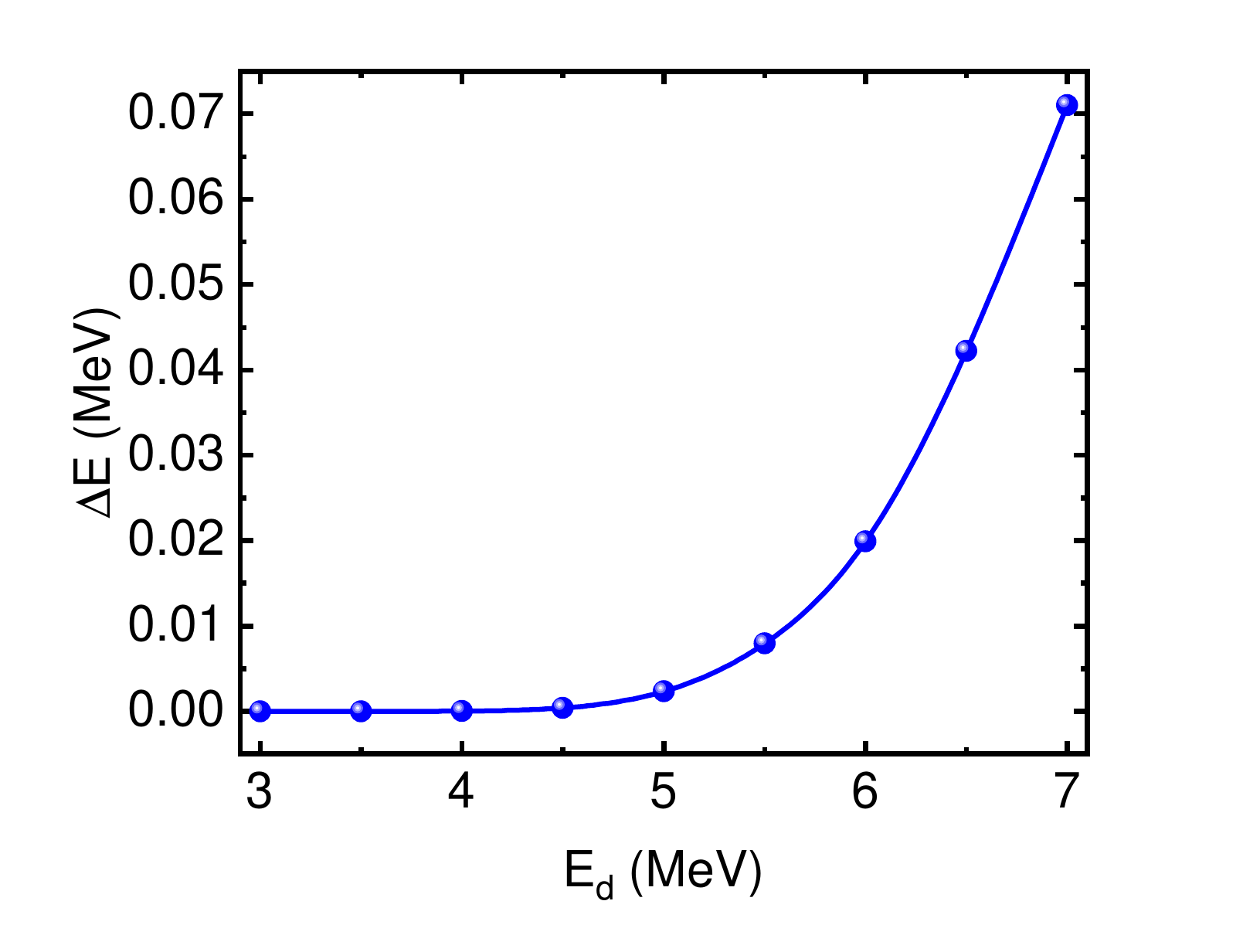}
\end{center}
\caption{(Color online) Average excitation energy as a function of the incident energy for d+$^{208}$Pb scattering with the scattering angle $\theta=150^\circ$ at $t=+\infty$.
%$R(E_d)$ is calculated by the tBF method employing different trajectories. The blue dotted curve is obtained with the Rutherford trajectory and the red solid curve is calculated by the modified Rutherford trajectory (with the correction of the $E1$ polarization potential). The experimental data (black solid dots with error bars) are also shown for comparison.
}\label{fig5}
\end{figure}

The semiclassical tBF approach adopted in this work neglects the effect of the energy loss on the center of mass motion of the projectile. The energy transferred to the intrinsic degree of freedom of the projectile should be compensated by a reduction of the energy in its center of mass degree of freedom. In Fig. \ref{fig5} we show the average excitation energy, which we refer to as $\Delta E$, after the scattering of the deuteron projectile on $^{208}$Pb target ($t=+\infty$) as a function of the incident energy at the maximum scattering angle ($\theta=150^\circ$) considered in this work. We find in Fig. \ref{fig5} that the average excitation energy increases significantly with the incident energy though its maximum value (at $E_d=7$ MeV) is only about $0.071$ MeV.

We evaluate the correction of the energy loss to the classical cross section [$\left(\frac{d\sigma}{d\Omega}\right)_{\rm class}$ in Eq.~(\ref{eq:CSection})] by an approximate method in Refs.~\cite{Alder:1956im,Boer:1977}
 \begin{eqnarray}
\left(\frac{d\sigma}{d\Omega}\right)^{EL}_{\rm class}=\frac{E_d}{E_d-\Delta E}\left(\frac{d\sigma}{d\Omega}\right)_{\rm class},
\label{eq:enerloss}
\end{eqnarray}
where $\left(\frac{d\sigma}{d\Omega}\right)^{EL}_{\rm class}$ represents the classical cross section with the correction of the energy loss. Accordingly, we introduce an uncertainty band (light gray region) for the dark green solid line in panel (a) of Fig. \ref{fig4} considering the correction of the energy loss to $R(E_d)$. It should be noted that the energy loss merely leads to the increase of $R(E_d)$. Therefore there is no uncertainty band induced by the energy loss below the dark green solid line. We observe in panel (a) of Fig. \ref{fig4} that the uncertainty induced by the energy loss is negligible at low incident energies while it increases noticeably with energies. In particular, the energy loss of about $0.071$ MeV leads to about $1\%$ uncertainty of $R(E_d)$ at $E_d=7$ MeV, which is much larger than the experimental uncertainty. Therefore, in order to address scattering more accurately with increasing projectile energy, we will improve the tBF method to describe the center of mass motion and the internal motion of the projectile in a self-consistent manner in the future.

\begin{figure}[tbh]
\begin{center}
\includegraphics[width=0.8\textwidth]{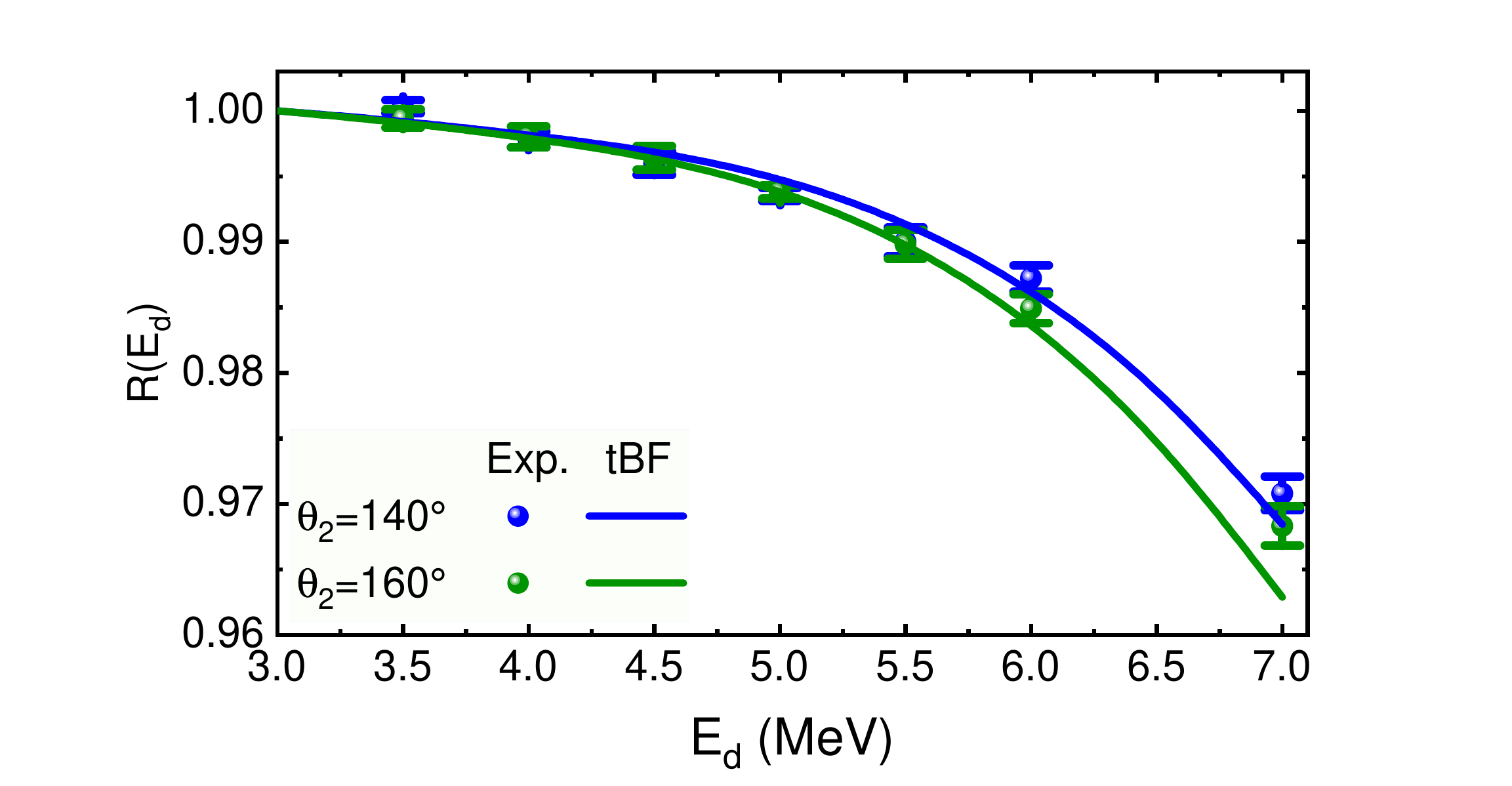}
\end{center}
\caption{(Color online) Elastic cross section ratios $R(E_d)$ with $(\theta_1,\theta_2)=(60^\circ,140^\circ)$ (blue color) and $(\theta_1,\theta_2)=(60^\circ,160^\circ)$ (green color) as functions of the bombarding energy $E_d$. The solid curves represent the tBF results. The solid dots with error bars denote the experimental data~\cite{Rodning:1982zz}.
}\label{fig6}
\end{figure}
Similarly, in Fig. \ref{fig6} we show $R(E_d)$ with $(\theta_1,\theta_2)=(60^\circ,140^\circ)$ and $(60^\circ,160^\circ)$ [see Eq.~(\ref{eq:RE})] for the scattering of the deuteron on $^{208}$Pb at $E_d=3-7$ MeV calculated by the tBF method and compared with experiment. We omit $\theta_1$ in the following since we use the same $\theta_1=60^\circ$. In the tBF calculations we take into account the correction of the polarization potential to the Rutherford trajectory, where we use {\it E1} polarizability $\alpha=0.635$ fm$^3$ (the converged value in Fig. \ref{fig_1}). We find in Fig. \ref{fig6} that the tBF method is able to reproduce the experimental $R(E_d)$ with $\theta_2=140^\circ$ and $160^\circ$ at $E_d=3-6$ MeV while falling below experiment at $E_d=7$ MeV. As we observe in Fig. \ref{fig4}, the discrepancy between the tBF results and the experimental data at $E_d=7$ MeV is due to the omission of the correction of the energy loss to the center of mass motion of the projectile in the tBF method. Especially, combining with the results for $\theta_2=150^\circ$ in Fig. \ref{fig4}, the deviation of the tBF results from the experimental data at $7$ MeV increases with $\theta_2$ since the energy loss increases monotonically with the scattering angle.

\section{Summary and Conclusions}
\label{sec:conclusions}
We investigated the scattering of the deuteron projectile on the $^{208}$Pb target below the Coulomb barrier based on the non-perturbative time-dependent basis function (tBF) approach.  We constructed the basis representation of the deuteron ground state and discretized scattering states of the {\it np} system by diagonalizing a realistic Hamiltonian based on the LENPIC {\it NN} interaction at N$^4$LO in a sufficiently large harmonic oscillator basis. In our calculations, we employed the {\it E1} polarizability $\alpha$ (in the polarization potential) obtained with the same {\it NN} interaction and consistent with the two existing experimental values.  We then applied the non-perturbative tBF approach to take higher-order {\it E1} transitions into account. We showed significant higher-order effects were present by comparing the populations of {\it E1} allowed and forbidden states after a scattering of d+$^{208}$Pb at $E_d=7$ MeV and $\theta=150^\circ$. We further noticed in the same scattering process that the contribution of the higher-order transitions to the inelastic channels was more than two orders of magnitude larger than that of the {\it E1} allowed couplings alone. By considering all the possible {\it E1} transition paths among all the states involved in the tBF approach and taking into account the corrections of the polarization potential to Rutherford trajectories, we successfully reproduced the quantity $R(E_d)$ with $(\theta_1,\theta_2)=(60^\circ,150^\circ)$ measured in experiment for $3$ MeV $<E_d<7$ MeV~\cite{Rodning:1982zz}. We found that both the internal {\it E1} transitions of the deuteron projectile and the corrections of the polarization potential to the classical Rutherford trajectories were essential for reproducing experimental data in these sub barrier experiments. More specifically, the correction of the polarization potential to the Rutherford trajectory played the dominant role in reproducing experimental data at the lowest bombarding energies that we considered while the role of the internal {\it E1} transitions of the deuteron projectile became increasingly significant as the bombarding energy increased and was dominant at $E_d=7$ MeV. We found the polarization potential provided a factor of $\sim12$ improvement in $\chi^2$/data compared to the result with no polarization potential. Finally, we also reproduced the experimental $R(E_d)$ with $(\theta_1,\theta_2)=(60^\circ,140^\circ)$ and $(\theta_1,\theta_2)=(60^\circ,160^\circ)$ at $3$ MeV $<E_d<7$ MeV~\cite{Rodning:1982zz}. We obtained all of our results without any adjustable parameters.

It should be noted that we have introduced a semiclassical approximation in the current tBF method, where the center of mass motion of the projectile is calculated via classical mechanics. We will extend the tBF method to a fully quantum mechanical framework in the future. In the current work, we considered the contributions of the {\it E1} effects separately for the deuteron's internal and center of mass degrees of freedom. The effect of the {\it E1} transitions in the internal degree of freedom of the deuteron projectile was taken into account by solving the time-dependent Schr\"{o}dinger equation. The effect of the {\it E1} polarization on the center of mass motion of the projectile was approximated by a polarization potential. In addition, we neglected the effect of the energy loss (due to the internal transitions) of the projectile on its center of mass motion. We found that the correction of the energy loss to the center of mass motion was negligible at the lower incident energies of the available data while its effects on $R(E_d)$ increased significantly with the incident energy, which led to underestimation of the tBF calculations to the experimental $R(E_d)$ at $E_d=7$ MeV. Within the framework of the fully quantum mechanical calculation, which remains for a future work, the center of mass motion and the internal motion of the projectile will be considered coherently.

\section*{Acknowledgments}
We acknowledge helpful discussions with Andrey Shirokov, Pieter Maris, Antonio M. Moro, Gerhard Baur, Zhigang Xiao, Li Ou, William Lynch and Betty Tsang. This work was supported in part by the US Department of Energy (DOE) under Grant Nos. DE-FG02-87ER40371 and DE-SC00018223. A portion of the computational resources were provided by the National Energy Research Scientific Computing Center (NERSC), which is supported by the US DOE Office of Science. Xingbo Zhao was supported by new faculty startup funding from the Institute of Modern Physics, Chinese Academy of Sciences, by Key Research Program of Frontier Sciences, CAS, Grant No. ZDBS-LY-7020 and by the Funds for Creative Research Groups of Gansu Province, China, Grant No. 20JR10RA067. Peng Yin and Wei Zuo were supported by the National Natural Science Foundation of China (Grant Nos. 11975282, 11705240, 11435014), the Strategic Priority Research Program of Chinese Academy of Sciences, Grant No. XDB34000000 and the Key Research Program of the Chinese Academy of Sciences (Grant No. XDPB15). This work was also partially supported by the CUSTIPEN (China-U.S. Theory Institute for Physics with Exotic Nuclei) funded by the U.S. Department of Energy, office of Science under Grant No. DE-SC0009971.

\end{document}